\documentclass[conference,a4paper]{IEEEtran}

\usepackage[english]{babel}
\usepackage[utf8]{inputenc}
\usepackage[T1]{fontenc}
\usepackage{hyphenat}

\usepackage{times}
\usepackage{amsmath,amssymb,amstext,amsfonts,mathrsfs,amsthm,mathtools,cases}
\usepackage{comment}

\usepackage{graphicx}                           
\graphicspath{{.}{./Figures/}}
\usepackage[tight]{subfigure}                   

\usepackage[nolist,printonlyused]{acronym}      

\usepackage[sort]{cite}                         

\usepackage{enumerate}                          

\usepackage{multirow}
\usepackage{tabularx}
\usepackage{booktabs}
\newcolumntype{C}{>{\centering\arraybackslash}X}
\newcolumntype{R}{>{\flushright\arraybackslash}X}
\newcolumntype{L}{>{\flushleft\arraybackslash}X}

\usepackage{algorithmic}
\usepackage[ruled]{algorithm}
\algsetup{indent=2ex}
\usepackage{eqparbox}

\usepackage{setspace}
\usepackage{xspace} 
\newcommand*{\unit}[2]{\mbox{\ensuremath{#1\,\mathrm{#2}}}}

\usepackage{float}
\floatname{algorithm}{\footnotesize Algorithm}

\usepackage{url}

\usepackage[normalem]{ulem}

\newcommand*{\figref}[1]{Figure~\ref{#1}}

\newcommand*{\tabref}[1]{Table~\ref{#1}}
\newcommand*{\algref}[1]{Algorithm~\ref{#1}}

\theoremstyle{definition}

\newtheorem{prop}{Proposition}

\DeclareMathOperator*{\argmin}{arg\,min}

\usepackage{color}

\newcommand{\comm}[1]{}

\DeclareMathAlphabet{\mathppl}{T1}{ppl}{m}{it}
\DeclareMathAlphabet{\mathphv}{T1}{phv}{m}{it}
\DeclareMathAlphabet{\mathpzc}{T1}{pzc}{m}{it}
\newlength{\norlen} 

\DeclareMathOperator{\diag}{diag}

\DeclareMathOperator{\re}{Re}

\DeclareMathOperator{\tr}{tr}

\newcommand{\Abs}[1]{\left\vert #1 \right\vert}

\newcommand{\Diag}[1]{\mathrm{Diag}\left( #1 \right)}

\newcommand{\Field}[1]{\mathbb{\uppercase{#1}}}

\newcommand{\Herm}[1]{{#1}^{\mathrm{H}}}

\newcommand{\Inv}[1]{{#1}^{-1}}

\newcommand{\Mt}[1]{\mathbf{#1}}
\newcommand{\Norm}[1]{\left\Vert #1 \right\Vert}

\newcommand{\NormTwo}[1]{\Norm{#1}_{2}}

\newcommand{\Real}[1]{\re\left\{#1\right\}}

\newcommand{\Set}[1]{\mathcal{\uppercase{#1}}}

\newcommand{\Trace}[1]{\tr\left(#1\right)}
\newcommand{\Transp}[1]{{#1}^{\mathrm{T}}}

\newcommand{\Vt}[1]{\mathbf{\lowercase{#1}}}
\newcommand{\mtA}{\Mt{A}}
\newcommand{\mtB}{\Mt{B}}

\newcommand{\mtH}{\Mt{H}}
\newcommand{\mtI}{\Mt{I}}

\newcommand{\mtR}{\Mt{R}}

\newcommand{\mtW}{\Mt{W}}
\newcommand{\mtX}{\Mt{X}}

\newcommand{\mtLambda}{\Mt{\Lambda}}
\newcommand{\mtPhi}{\Mt{\Phi}}
\newcommand{\mtPsi}{\Mt{\Psi}}
\newcommand{\mtSigma}{\Mt{\Sigma}}
\newcommand{\mtGamma}{\Mt{\Gamma}}

\newcommand{\mtTheta}{\Mt{\Theta}}

\newcommand{\vtA}{\Vt{A}}
\newcommand{\vtB}{\Vt{B}}
\newcommand{\vtC}{\Vt{C}}

\newcommand{\vtE}{\Vt{E}}

\newcommand{\vtH}{\Vt{H}}

\newcommand{\vtR}{\Vt{R}}
\newcommand{\vtS}{\Vt{S}}

\newcommand{\vtW}{\Vt{W}}
\newcommand{\vtX}{\Vt{X}}
\newcommand{\vtY}{\Vt{Y}}
\newcommand{\vtZ}{\Vt{Z}}

\newcommand{\vtEta}{\Vt{\boldsymbol{\eta}}}

\newcommand{\vtMu}{\Vt{\boldsymbol{\mu}}}

\newcommand{\vtOne}{\Vt{1}}

\newcommand{\vtZero}{\Vt{0}}

\newcommand{\fdC}{\Field{C}}

\newcommand{\fdR}{\Field{R}}

\newcommand{\stC}{\Set{C}}

\newcommand{\stN}{\Set{N}}

\newcommand{\bbE}{\mathbb{E}}

\IEEEoverridecommandlockouts

\allowdisplaybreaks[4]

\begin{document}
\date{\today}

\bstctlcite{IEEEexample:BSTcontrol}

\title{How to Split UL/DL Antennas in Full-Duplex Cellular Networks}

\author{
Jos\'{e} Mairton B.~da Silva Jr$^{\star}$,
\thanks{Jos\'{e} Mairton B. da Silva Jr. would like to acknowledge CNPq, a Brazilian
research-support agency. The simulations were performed on resources provided by the Swedish
National Infrastructure for Computing (SNIC) at PDC Centre for High Performance Computing
(PDC-HPC). G. Fodor was partially supported by the Wireless@KTH project Naomi.}
Hadi Ghauch$^{\star}$, G\'{a}bor Fodor$^{\star\dagger}$, Carlo Fischione${^\star}$\\
$^\star$KTH, Royal Institute of Technology, Stockholm, Sweden \\
$^\dagger$Ericsson Research, Kista, Sweden
\\[0pt]\vspace{-0.95\baselineskip}
}


\begin{acronym}[LTE-Advanced]
  \acro{2G}{Second Generation}
  \acro{3-DAP}{3-Dimensional Assignment Problem}
  \acro{3G}{3$^\text{rd}$~Generation}
  \acro{3GPP}{3$^\text{rd}$~Generation Partnership Project}
  \acro{4G}{4$^\text{th}$~Generation}
  \acro{5G}{5$^\text{th}$~Generation}
  \acro{AA}{Antenna Array}
  \acro{AC}{Admission Control}
  \acro{AD}{Attack-Decay}
  \acro{ADC}{analog-to-digital converter}
  \acro{ADSL}{Asymmetric Digital Subscriber Line}
  \acro{AHW}{Alternate Hop-and-Wait}
  \acro{AMC}{Adaptive Modulation and Coding}
  \acro{AP}{Access Point}
  \acro{APA}{Adaptive Power Allocation}
  \acro{ARMA}{Autoregressive Moving Average}
  \acro{ATES}{Adaptive Throughput-based Efficiency-Satisfaction Trade-Off}
  \acro{AWGN}{Additive White Gaussian Noise}
  \acro{BB}{Branch and Bound}
  \acro{BCD}{block coordinate descent}
  \acro{BD}{Block Diagonalization}
  \acro{BER}{Bit Error Rate}
  \acro{BF}{Best Fit}
  \acro{BFD}{bidirectional full duplex}
  \acro{BLER}{BLock Error Rate}
  \acro{BPC}{Binary Power Control}
  \acro{BPSK}{Binary Phase-Shift Keying}
  \acro{BRA}{Balanced Random Allocation}
  \acro{BS}{base station}
  \acro{BSUM}{block successive upper-bound minimization}
  \acro{CAP}{Combinatorial Allocation Problem}
  \acro{CAPEX}{Capital Expenditure}
  \acro{CBF}{Coordinated Beamforming}
  \acro{CBR}{Constant Bit Rate}
  \acro{CBS}{Class Based Scheduling}
  \acro{CC}{Congestion Control}
  \acro{CDF}{cumulative distribution function}
  \acro{CDMA}{Code-Division Multiple Access}
  \acro{CL}{Closed Loop}
  \acro{CLPC}{Closed Loop Power Control}
  \acro{CNR}{Channel-to-Noise Ratio}
  \acro{CPA}{Cellular Protection Algorithm}
  \acro{CPICH}{Common Pilot Channel}
  \acro{CoMP}{Coordinated Multi-Point}
  \acro{CQI}{Channel Quality Indicator}
  \acro{CRM}{Constrained Rate Maximization}
	\acro{CRN}{Cognitive Radio Network}
  \acro{CS}{Coordinated Scheduling}
  \acro{CSI}{channel state information}
  \acro{CUE}{Cellular User Equipment}
  \acro{D2D}{Device-to-Device}
  \acro{DAC}{digital-to-analog converter}
  \acro{DCA}{Dynamic Channel Allocation}
  \acro{DE}{Differential Evolution}
  \acro{DFT}{Discrete Fourier Transform}
  \acro{DIST}{Distance}
  \acro{DL}{downlink}
  \acro{DMA}{Double Moving Average}
	\acro{DMRS}{Demodulation Reference Signal}
  \acro{D2DM}{D2D Mode}
  \acro{DMS}{D2D Mode Selection}
  \acro{DPC}{Dirty Paper Coding}
  \acro{DRA}{Dynamic Resource Assignment}
  \acro{DSA}{Dynamic Spectrum Access}
  \acro{DSM}{Delay-based Satisfaction Maximization}
  \acro{ECC}{Electronic Communications Committee}
  \acro{EFLC}{Error Feedback Based Load Control}
  \acro{EI}{Efficiency Indicator}
  \acro{eNB}{Evolved Node B}
  \acro{EPA}{Equal Power Allocation}
  \acro{EPC}{Evolved Packet Core}
  \acro{EPS}{Evolved Packet System}
  \acro{E-UTRAN}{Evolved Universal Terrestrial Radio Access Network}
  \acro{ES}{Exhaustive Search}
  \acro{FD}{full-duplex}
  \acro{FDD}{frequency division duplex}
  \acro{FDM}{Frequency Division Multiplexing}
  \acro{FER}{Frame Erasure Rate}
  \acro{FF}{Fast Fading}
  \acro{FSB}{Fixed Switched Beamforming}
  \acro{FST}{Fixed SNR Target}
  \acro{FTP}{File Transfer Protocol}
  \acro{GA}{Genetic Algorithm}
  \acro{GBR}{Guaranteed Bit Rate}
  \acro{GLR}{Gain to Leakage Ratio}
  \acro{GOS}{Generated Orthogonal Sequence}
  \acro{GPL}{GNU General Public License}
  \acro{GRP}{Grouping}
  \acro{HARQ}{Hybrid Automatic Repeat Request}
  \acro{HD}{half-duplex}
  \acro{HMS}{Harmonic Mode Selection}
  \acro{HOL}{Head Of Line}
  \acro{HSDPA}{High-Speed Downlink Packet Access}
  \acro{HSPA}{High Speed Packet Access}
  \acro{HTTP}{HyperText Transfer Protocol}
  \acro{ICMP}{Internet Control Message Protocol}
  \acro{ICI}{Intercell Interference}
  \acro{ID}{Identification}
  \acro{IETF}{Internet Engineering Task Force}
  \acro{ILP}{Integer Linear Program}
  \acro{JRAPAP}{Joint RB Assignment and Power Allocation Problem}
  \acro{UID}{Unique Identification}
  \acro{IID}{Independent and Identically Distributed}
  \acro{IIR}{Infinite Impulse Response}
  \acro{ILP}{Integer Linear Problem}
  \acro{IMT}{International Mobile Telecommunications}
  \acro{INV}{Inverted Norm-based Grouping}
	\acro{IoT}{Internet of Things}
  \acro{IP}{Integer Programming}
  \acro{IPv6}{Internet Protocol Version 6}
  \acro{ISD}{Inter-Site Distance}
  \acro{ISI}{Inter Symbol Interference}
  \acro{ITU}{International Telecommunication Union}
  \acro{JAFM}{joint assignment and fairness maximization}
  \acro{JAFMA}{joint assignment and fairness maximization algorithm}
  \acro{JOAS}{Joint Opportunistic Assignment and Scheduling}
  \acro{JOS}{Joint Opportunistic Scheduling}
  \acro{JP}{Joint Processing}
	\acro{JS}{Jump-Stay}
  \acro{KKT}{Karush-Kuhn-Tucker}
  \acro{L3}{Layer-3}
  \acro{LAC}{Link Admission Control}
  \acro{LA}{Link Adaptation}
  \acro{LC}{Load Control}
  \acro{LOS}{line of sight}
  \acro{LP}{Linear Programming}
  \acro{LTE}{Long Term Evolution}
	\acro{LTE-A}{\ac{LTE}-Advanced}
  \acro{LTE-Advanced}{Long Term Evolution Advanced}
  \acro{M2M}{Machine-to-Machine}
  \acro{MAC}{medium access control}
  \acro{MANET}{Mobile Ad hoc Network}
  \acro{MC}{Modular Clock}
  \acro{MCS}{Modulation and Coding Scheme}
  \acro{MDB}{Measured Delay Based}
  \acro{MDI}{Minimum D2D Interference}
  \acro{MF}{Matched Filter}
  \acro{MG}{Maximum Gain}
  \acro{MH}{Multi-Hop}
  \acro{MIMO}{multiple input multiple output}
  \acro{MINLP}{mixed integer nonlinear programming}
  \acro{MIP}{Mixed Integer Programming}
  \acro{MISO}{multiple input single output}
  \acro{MLWDF}{Modified Largest Weighted Delay First}
  \acro{MME}{Mobility Management Entity}
  \acro{MMSE}{minimum mean squared error}
  \acro{MOS}{Mean Opinion Score}
  \acro{MPF}{Multicarrier Proportional Fair}
  \acro{MRA}{Maximum Rate Allocation}
  \acro{MR}{Maximum Rate}
  \acro{MRC}{Maximum Ratio Combining}
  \acro{MRT}{Maximum Ratio Transmission}
  \acro{MRUS}{Maximum Rate with User Satisfaction}
  \acro{MS}{Mode Selection}
  \acro{MSE}{mean squared error}
  \acro{MSI}{Multi-Stream Interference}
  \acro{MTC}{Machine-Type Communication}
  \acro{MTSI}{Multimedia Telephony Services over IMS}
  \acro{MTSM}{Modified Throughput-based Satisfaction Maximization}
  \acro{MU-MIMO}{Multi-User Multiple Input Multiple Output}
  \acro{MU}{Multi-User}
  \acro{NAS}{Non-Access Stratum}
  \acro{NB}{Node B}
	\acro{NCL}{Neighbor Cell List}
  \acro{NLP}{Nonlinear Programming}
  \acro{NLOS}{non-line of sight}
  \acro{NMSE}{Normalized Mean Square Error}
  \acro{NORM}{Normalized Projection-based Grouping}
  \acro{NP}{non-polynomial time}
  \acro{NRT}{Non-Real Time}
  \acro{NSPS}{National Security and Public Safety Services}
  \acro{O2I}{Outdoor to Indoor}
  \acro{OFDMA}{Orthogonal Frequency Division Multiple Access}
  \acro{OFDM}{Orthogonal Frequency Division Multiplexing}
  \acro{OFPC}{Open Loop with Fractional Path Loss Compensation}
	\acro{O2I}{Outdoor-to-Indoor}
  \acro{OL}{Open Loop}
  \acro{OLPC}{Open-Loop Power Control}
  \acro{OL-PC}{Open-Loop Power Control}
  \acro{OPEX}{Operational Expenditure}
  \acro{ORB}{Orthogonal Random Beamforming}
  \acro{JO-PF}{Joint Opportunistic Proportional Fair}
  \acro{OSI}{Open Systems Interconnection}
  \acro{PAIR}{D2D Pair Gain-based Grouping}
  \acro{PAPR}{Peak-to-Average Power Ratio}
  \acro{P2P}{Peer-to-Peer}
  \acro{PC}{Power Control}
  \acro{PCI}{Physical Cell ID}
  \acro{PDCCH}{physical downlink control channel}
  \acro{PDF}{Probability Density Function}
  \acro{PER}{Packet Error Rate}
  \acro{PF}{Proportional Fair}
  \acro{P-GW}{Packet Data Network Gateway}
  \acro{PL}{Pathloss}
  \acro{PRB}{Physical Resource Block}
  \acro{PROJ}{Projection-based Grouping}
  \acro{ProSe}{Proximity Services}
  \acro{PS}{Packet Scheduling}
  \acro{PSCA}{parallel successive convex approximation}
  \acro{PSO}{Particle Swarm Optimization}
  \acro{PUCCH}{physical uplink control channel}
  \acro{PZF}{Projected Zero-Forcing}
  \acro{QAM}{Quadrature Amplitude Modulation}
  \acro{QoS}{quality of service}
  \acro{QPSK}{Quadri-Phase Shift Keying}
  \acro{RAISES}{Reallocation-based Assignment for Improved Spectral Efficiency and Satisfaction}
  \acro{RAN}{Radio Access Network}
  \acro{RA}{Resource Allocation}
  \acro{RAT}{Radio Access Technology}
  \acro{RATE}{Rate-based}
  \acro{RB}{resource block}
  \acro{RBG}{Resource Block Group}
  \acro{REF}{Reference Grouping}
  \acro{RF}{Radio-Frequency}
  \acro{RLC}{Radio Link Control}
  \acro{RM}{Rate Maximization}
  \acro{RNC}{Radio Network Controller}
  \acro{RND}{Random Grouping}
  \acro{RRA}{Radio Resource Allocation}
  \acro{RRM}{Radio Resource Management}
  \acro{RSCP}{Received Signal Code Power}
  \acro{RSRP}{Reference Signal Receive Power}
  \acro{RSRQ}{Reference Signal Receive Quality}
  \acro{RR}{Round Robin}
  \acro{RRC}{Radio Resource Control}
  \acro{RSSI}{Received Signal Strength Indicator}
  \acro{RT}{Real Time}
  \acro{RU}{Resource Unit}
  \acro{RUNE}{RUdimentary Network Emulator}
  \acro{RV}{Random Variable}
  \acro{SAC}{Session Admission Control}
  \acro{SCM}{Spatial Channel Model}
  \acro{SC-FDMA}{Single Carrier - Frequency Division Multiple Access}
  \acro{SCP}{sequential convex programming}
  \acro{SD}{Soft Dropping}
  \acro{S-D}{Source-Destination}
  \acro{SDPC}{Soft Dropping Power Control}
  \acro{SDMA}{Space-Division Multiple Access}
  \acro{SDR}{semidefinite relaxation}
  \acro{SDP}{semidefinite programming}
  \acro{SER}{Symbol Error Rate}
  \acro{SES}{Simple Exponential Smoothing}
  \acro{S-GW}{Serving Gateway}
  \acro{SINR}{signal-to-interference-plus-noise ratio}
  \acro{SI}{self-interference}
  \acro{SIP}{Session Initiation Protocol}
  \acro{SISO}{Single Input Single Output}
  \acro{SIMO}{Single Input Multiple Output}
  \acro{SIR}{Signal to Interference Ratio}
  \acro{SLNR}{Signal-to-Leakage-plus-Noise Ratio}
  \acro{SMA}{Simple Moving Average}
  \acro{SNR}{signal to noise ratio}
  \acro{SORA}{Satisfaction Oriented Resource Allocation}
  \acro{SORA-NRT}{Satisfaction-Oriented Resource Allocation for Non-Real Time Services}
  \acro{SORA-RT}{Satisfaction-Oriented Resource Allocation for Real Time Services}
  \acro{SPF}{Single-Carrier Proportional Fair}
  \acro{SRA}{Sequential Removal Algorithm}
  \acro{SRS}{Sounding Reference Signal}
  \acro{SU-MIMO}{Single-User Multiple Input Multiple Output}
  \acro{SU}{Single-User}
  \acro{SVD}{Singular Value Decomposition}
  \acro{TCP}{Transmission Control Protocol}
  \acro{TDD}{time division duplex}
  \acro{TDMA}{Time Division Multiple Access}
  \acro{TNFD}{three node full duplex}
  \acro{TETRA}{Terrestrial Trunked Radio}
  \acro{TP}{Transmit Power}
  \acro{TPC}{Transmit Power Control}
  \acro{TTI}{Transmission Time Interval}
  \acro{TTR}{Time-To-Rendezvous}
  \acro{TSM}{Throughput-based Satisfaction Maximization}
  \acro{TU}{Typical Urban}
  \acro{UE}{user equipment}
  \acro{UEPS}{Urgency and Efficiency-based Packet Scheduling}
  \acro{UL}{uplink}
  \acro{UMTS}{Universal Mobile Telecommunications System}
  \acro{URI}{Uniform Resource Identifier}
  \acro{URM}{Unconstrained Rate Maximization}
  \acro{VR}{Virtual Resource}
  \acro{VoIP}{Voice over IP}
  \acro{WAN}{Wireless Access Network}
  \acro{WCDMA}{Wideband Code Division Multiple Access}
  \acro{WF}{Water-filling}
  \acro{WiMAX}{Worldwide Interoperability for Microwave Access}
  \acro{WINNER}{Wireless World Initiative New Radio}
  \acro{WLAN}{Wireless Local Area Network}
  \acro{WMMSE}{weighted minimum mean square error}
  \acro{WMPF}{Weighted Multicarrier Proportional Fair}
  \acro{WPF}{Weighted Proportional Fair}
  \acro{WSN}{Wireless Sensor Network}
  \acro{WWW}{World Wide Web}
  \acro{XIXO}{(Single or Multiple) Input (Single or Multiple) Output}
  \acro{ZF}{Zero-Forcing}
  \acro{ZMCSCG}{Zero Mean Circularly Symmetric Complex Gaussian}
\end{acronym}

\maketitle
\IEEEpeerreviewmaketitle

\newcommand*{\BS}[1]{\ensuremath{\text{BS}_{#1}}}
\newcommand*{\UE}[1]{\ensuremath{\text{UE}_{#1}}}

\begin{abstract}
To further improve the potential of full-duplex communications, networks may employ multiple
antennas at the base station or user equipment. 
To this end, networks that employ current
radios usually deal with self-interference and multi-user interference by beamforming techniques.
Although previous works investigated beamforming design to improve spectral efficiency, the
fundamental question of how to split the antennas at a base station between uplink and downlink in 
full-duplex networks has not been investigated rigorously.
This paper addresses this question by posing antenna splitting as a binary nonlinear optimization 
problem to minimize the sum mean squared error of the received data symbols. It is shown that this 
is an NP-hard problem.
This combinatorial problem is dealt with by equivalent formulations, iterative
convex approximations, and a binary relaxation.
The proposed algorithm is guaranteed to converge to a stationary solution of the relaxed problem
with much smaller complexity than exhaustive search.
Numerical results indicate that the proposed solution is close to the optimal in both high and low
self-interference capable scenarios, while the usually assumed antenna splitting is far from
optimal. 
For large number of antennas, a simple antenna splitting is close to the proposed
solution. 
This reveals that the importance of antenna splitting diminishes with the number of antennas.
\end{abstract}


\section{Introduction}\label{sec:intro}

Traditional cellular networks operate in \ac{HD} transmission mode, in which a \ac{UE} and the
\ac{BS} either transmits or receives on a given frequency channel. 
Due to recent advancements in
antenna and radio-frequency/analog interference cancellation techniques, 
\ac{FD} transmissions appear as a viable alternative to traditional \ac{HD} transmission
modes~\cite{Sabharwal2014}.
\ac{FD} transmission mode overcomes the assumption that it is not possible for radios
to transmit and receive simultaneously on the same time-frequency resource, and can almost double
the spectral efficiency of conventional \ac{HD} wireless transmission modes, especially in the low
transmit power domain~\cite{Sabharwal2014,Goyal2015,Mairton2017}.

Due to the developments in \ac{FD} communications, \ac{MIMO} techniques at the \ac{BS} or
\acp{UE} are becoming a realistic technology component of advanced wireless systems, based on both
theoretical~\cite{Day2012,Nguyen2014,Cirik2016,Zhou2014} 
and practical~\cite{Bharadia2014,Everett2016,Gowda2017}
analyses. 
The employment of multiple antennas with \ac{FD} communications is possible using
current radio components that can either transmit or receive on the same time-frequency resource,
or specialized radio components equipped with a duplexer that allows antennas to transmit and 
receive on the same time-frequency resource~\cite{Sabharwal2014}.

\begin{figure}
\centering
\includegraphics[width=0.75\linewidth,trim=0mm 0mm 0mm 0mm,clip]{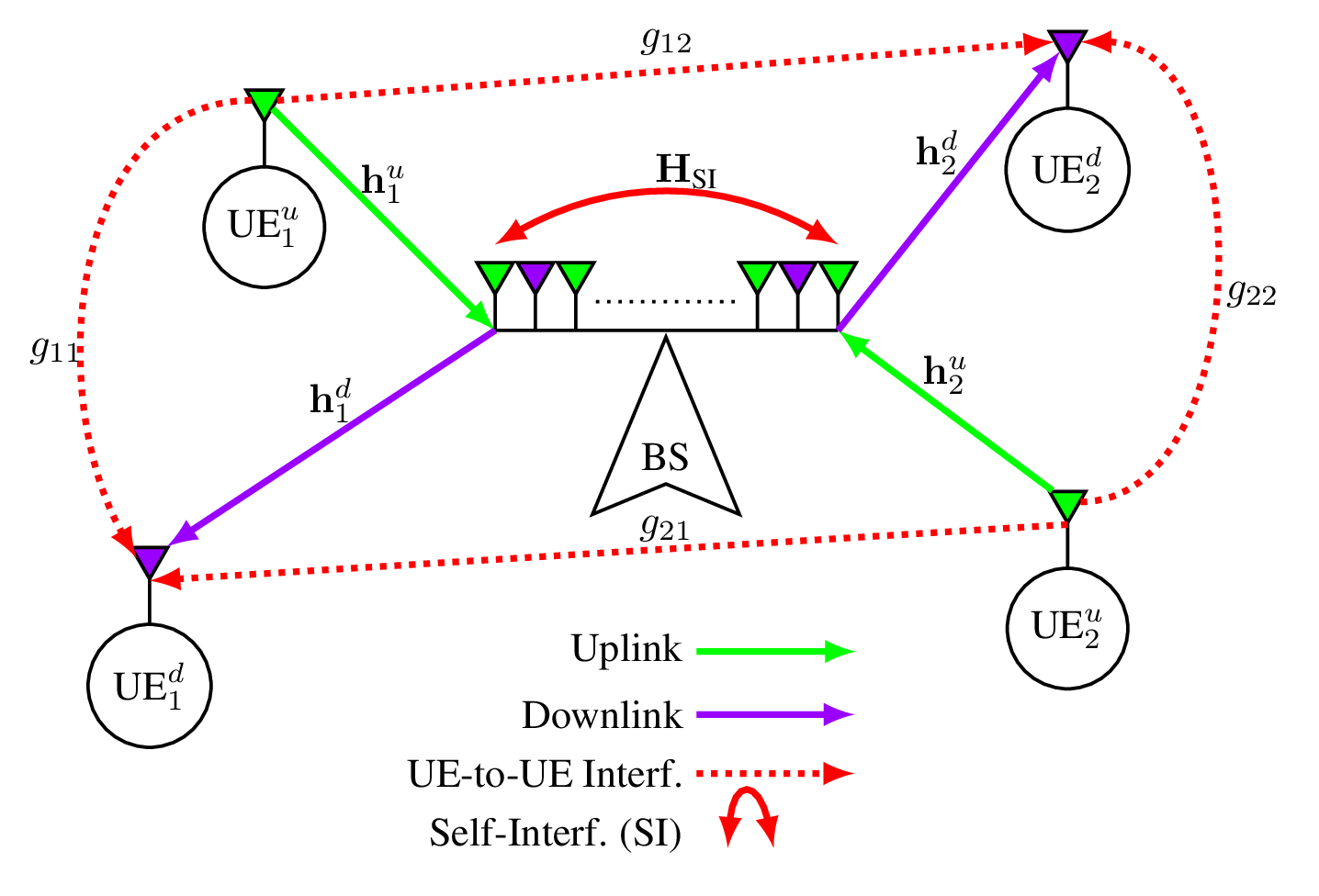}
\caption{An example of a multi-antenna cellular network employing \ac{FD} with two \ac{UE} pairs.
The \ac{BS} uses a subset of the antennas for \ac{UL} reception (green),
while the remaining (purple) antennas are used for \ac{DL} transmission.
To mitigate all interferences, it is necessary to analyse the impact of splitting the set of the
available antennas between \ac{UL} and \ac{DL}.}\label{fig:scenario_fd_mimo}
\end{figure}

An example of a cellular network employing such radios with \ac{MIMO} and two single-antenna
\acp{UE} pairs is illustrated in Figure~\ref{fig:scenario_fd_mimo}.
Note that apart from the inherently present \ac{SI} from the \ac{DL} (in purple) to the \ac{UL}
antennas (in green), \ac{FD} operation in a cellular network must also deal with the
\ac{UE}-to-\ac{UE} and multi-user interference, indicated by the red dotted lines between users.
To mitigate the negative effects of both interferences on the spectral efficiency of
the system, coordination mechanisms are needed~\cite{Sabharwal2014,Goyal2015}.
Specifically, a key element is the splitting of \ac{UL} and \ac{DL} antennas, which impacts the
number of available antennas for transmission and reception, as well as the characteristics of 
the self-interference channel, and the power level at which the \ac{UL} and \ac{DL} signals will 
be received.
Consequently, it is crucial to understand how the \ac{UL}/\ac{DL} antennas in multi-antenna
full-duplex cellular networks should be split.

Most of the works in \ac{FD} networks assume the antenna splitting between \ac{UL} and \ac{DL}
antennas is already given, and aim to analyse the theoretical improvement achieved by 
\ac{FD}~\cite{Day2012,Nguyen2014,Cirik2016}. 
Nevertheless, some works in the literature address the general topic of antenna
splitting~\cite{Zhou2014,Everett2016,Gowda2017}. The work in~\cite{Zhou2014} considers antenna
selection in a bidirectional \ac{FD} system with two antennas to maximize the sum rate or minimize 
the symbol error rate, when operating in a point-to-point single-antenna scenario. 
The authors
in~\cite{Everett2016} analyse \ac{SI} cancellation via digital beamforming for large-antennas
\ac{FD} communications, whose proposed solution highlights the importance of
\ac{UL}/\ac{DL} antenna splitting and assumes a fixed splitting. Similarly, the work
in~\cite{Gowda2017} devises antenna splitting and beamforming to minimize the gap between demand
and achievable rates. 
Assuming a given number of full-duplex antennas with analog cancellation
into the possible splits, and no \ac{UE}-to-\ac{UE} interference between single-antenna \ac{UL}
and \ac{DL} users, the proposed suboptimal algorithm splits the antennas and evaluates the
\ac{DL} beamforming to minimize the self-interference on receive antennas.

However, the fundamental problem of \ac{UL}/\ac{DL} antenna splitting in a multi-user setting,
considering all interferences and distortions between \ac{UL} and \ac{DL} users, has not been
studied.
Therefore, in this work we address this limitation by proposing a combinatorial optimization
problem aiming to minimize the total \ac{MSE} of the received data symbols. 
Our proposed optimization problem poses technical challenges that are markedly different from 
those investigated in previous works~\cite{Zhou2014,Everett2016,Gowda2017}.
Using the inverse relation between \ac{MSE} and rate~\cite{Christensen2008}, we develop an
original and new problem formulation based on the minimization of the sum \ac{MSE} that considers
the antenna assignment between \ac{UL} and \ac{DL}, and that exhibits high complexity.
Our solution approach relies on rewriting the sum \ac{MSE} as sum of quadratic and biquadratic 
terms of the assignment variables of \ac{UL} and \ac{DL}, and 
then resorting to a first-order approximation.
Since the approximated problem is combinatorial and NP-hard, we relax the
binary constraints to the hypercube, and use the framework of \ac{PSCA}~\cite{Razaviyayn2014} to
solve to the original problem. 
The numerical results indicate that the
proposed solution is close to the optimal solution, while simple antenna splitting usually
assumed is far from optimal. 
Moreover, results show that optimized antenna splitting is
crucial for scenarios with low \ac{SI} cancellation capability, whereas its importance diminishes
as the number of antennas increases.

\emph{Notation:}
Vectors and matrices are denoted by bold lower and upper case letters, respectively;
$\Herm{\mtA},\;\Transp{\mtA},\;\mtA^*$ represent the Hermitian, transpose, and complex conjugate 
of $\mtA$, respectively; $\Diag{\mtA}$ is the column vector created from the
diagonal of matrix $\mtA$; similarly, $\diag(\mtA)$ or $\diag(\vtA)$ are the diagonal matrices
whose elements are in the diagonal of matrix $\mtA$, or composed by vector $\vtA$ in the diagonal,
respectively.
We denote by $\mtI_{K}$ the identity matrix of dimension $K$, by $\mathbf{0}$ and $\mathbf{1}$ a
vector or matrix where all elements are zero or one, respectively, and by $\fdC$ the complex field.
We denote expectation by $\bbE\{\cdot\}$, the Hadamard product between matrices by $\mtA\!\odot\! 
\mtB$, statistical orthogonality by $\perp$, the circular complex Gaussian distribution with mean 
vector $\vtMu$ and covariance matrix $\mtSigma$ by $\stC\stN(\vtMu,\mtSigma)$.

\section{System Model and Problem Formulation}\label{sec:sys_mod}

\subsection{System Model}
We consider a single-cell cellular system in which the \ac{BS} is \ac{FD}
capable, while the \acp{UE} are \ac{HD} capable, as illustrated by
\figref{fig:scenario_fd_mimo}. The \ac{BS} is equipped with $M$ antennas, which can be used to
serve $I$ \ac{UL} and $J$ \ac{DL} single-antenna users.
In the figure, the \ac{BS} is subject to \ac{SI} from the \ac{DL} antennas, whereas the \acp{UE}
in the \ac{UL} ($\UE{1}^u$ and $\UE{2}^u$) cause \ac{UE}-to-\ac{UE} interference to co-scheduled
\acp{UE} in the \ac{DL} ($\UE{1}^d$ and $\UE{2}^d$).
We let $s_{i}^u$ and $s_{j}^d$ denote the transmitted data symbol in the \ac{UL} and \ac{DL},
respectively, where both are zero mean with unit power.
The transmitted power in the \ac{UL} is denoted by $q_{i}^u\in\fdR$, and the linear beamformer in
the \ac{DL} by $\vtW_{j}^d\in\fdC^{M\times 1}$.

Let $\vtH_{i}^u\in\fdC^{M\times 1}$ denote the complex channel vector comprising large scale
fading, shadowing, and path-loss between transmitter \ac{UE} $i$ and the \ac{BS},
$\vtH_{j}^d\in\fdC^{M\times 1}$ denote the channel vector between the \ac{BS} and receiving
\ac{UE} $j$, and $g_{ij}\in\fdC$ denote the interfering channel gain between the UL transmitter
\ac{UE} $i$ and the DL receiver \ac{UE} $j$. All channel elements $\vtH_{i}^u,\vtH_{j}^d$, and
$g_{ij}$ have an independent and identically distributed (i.i.d.) complex Gaussian distribution
with zero mean and unit variance.
Let $\mtH_{\text{SI}}\in\fdC^{M\times M}$ denote the \ac{SI} channel matrix
from the transmit antennas in \ac{DL} to the receive antennas in the \ac{UL}, which is modelled as
Rician fading~\cite{Duarte2012,Nguyen2014}.
Accordingly, $\mtH_{\text{SI}}\!\sim\!\stC\stN\left(\sqrt{\sigma_{\text{SI}}^2
K_r/(1+K_r)}\mathbf{1}_{M\times M},\left(\sigma_{\text{SI}}^2/(1+K_r)\right)\mtI_M\right)$, where
$K_r$ is the Rician factor and assumed to be 1, and $\sigma_{\text{SI}}^2$ represents the
\ac{SI} cancellation capability.

The signal received by the \ac{BS} in
the \ac{UL}, $\vtY^u\!\!\in\!\!\fdC^M$, and by \ac{DL} user $j$, $y^d_j\!\in\!\fdC$,
respectively, can be written as
\small
\begin{align}
\vtY^u\!\! &=\!\! \sum\nolimits_{i=1}^{I}\!\! \vtH^u_i(\! 
\sqrt{q_{i}^u}s_i^u+c_i^u) \!+\! \mtH_{\text{SI}}\!\left(\sum\nolimits_{j=1}^J\!\! \vtW_j^d s_j^d 
\!+\! \vtC^d\!\right)\! \!+\! \vtEta^u \!+\! \vtE^u\!,\;\hspace*{-0.2cm}\label{eq:UL_signal}\\
y_j^d\!\! &=\!\! \Herm{\vtH^d_j}\!\left(\!\sum\nolimits_{m=1}^{J}\!\!\! \vtW_m^ds_m^d \!+\! 
\vtC^d\!\right)\! \!+\! \sum\nolimits_{i=1}^I\! g_{ij}\!\left(\!\!\sqrt{q_{i}^u}s_i^u \!+\! 
c_i^u\!\right) \!\!+\!\! \eta^d_j \!\!+\!\! e^d_j,\hspace*{-0.1cm}\label{eq:DL_signal}
\end{align}\normalsize
where $\vtEta^{u}\!\sim\!\stC\stN\left(\mathbf{0}_M,\sigma^2\mtI_M \right)$ and
$\eta_j^d\!\sim\!\stC\stN\left(0,\sigma^2\right)$ are additive white Gaussian noise at the
\ac{BS} and at \ac{DL} user $j$, respectively. Notice that the second term in~\eqref{eq:UL_signal}
denotes the \ac{SI}, whereas the second term in Eq.~\eqref{eq:DL_signal}
denotes the \ac{UE}-to-\ac{UE} interference caused by \ac{UL} to \ac{DL} users.
Moreover, multi-user interference is also present, as seen in the first summation
of~\eqref{eq:UL_signal}-\eqref{eq:DL_signal}.

To account for non-ideal circuits in the limited dynamic range, we consider an additional additive
white Gaussian distortion signal at the transmitter and receiver~\cite{Day2012}, which are
modelled in the \ac{UL} as $c_j^u\in\fdC$ and $\vtE^u\in\fdC^{M\times 1}$, and in the \ac{DL} as
$\vtC^d\in\fdC^{M\times 1}$ and $e^d_j\in\fdC$ in the \ac{DL}.
Following previous works~\cite{Day2012,Cirik2016}, we define the transmitter distortions in the 
\ac{UL} as $c_i^u \!\!\sim\!\stC\stN\left(0,\kappa q_i^u \right),\; c_i^u\!\perp\! c_k^u|_{k\neq 
i},\; c_i^u\!\perp\! s_i^u$, and in the \ac{DL} as $\vtC^d 
\!\!\sim\!\stC\stN\left(\mathbf{0}_M,\kappa\sum_{j=1}^{J}\diag(\vtW_j^d\Herm{\vtW_j^d})
\right),\; \vtC^d\! \perp\! \vtS^d$,
where typically $\kappa\ll 1$. Furthermore, the receiver distortion is modelled in the \ac{UL} as 
$\vtE^u \!\!\sim\!\stC\stN\left(\mathbf{0}_M,\beta\diag(\mtPhi^u) \right),\; \vtE^u\! \perp\!
\vtY^u\!-\!\vtE^u$, and in the \ac{DL} as $e_j^d \!\!\sim\!\stC\stN\left(0,\beta\Phi^d \right),\; 
e_j^d \!\perp\! y_j^d\!-\!e_j^d$,
where typically $\beta \ll 1$; $\vtY^u-\vtE^u$ and $y_j^d-e_j^d$ are the receiver undistorted 
signal in the \ac{UL} and at \ac{DL} user $j$; $\mtPhi^u\!\!\in\!\!\fdC^{M\!\times\! M}$ and
$\Phi^d\!\!\in\!\!\fdR$ denote the covariance matrix and variance of the received undistorted
vector in the \ac{UL} and at \ac{DL} user $j$, whose expressions are omitted due to space
limitations.
Note that the model characterized by the distortions 
closely approximates the combined effects of power amplifier noise, non-linearities in the
analog-to-digital and digital-to-analog converters, as well as the oscillator phase noise in
practical hardware~\cite{Day2012}.

As illustrated in \figref{fig:scenario_fd_mimo}, the $M$ antennas at the \ac{BS} may transmit and
receive, but the direction in \ac{UL} or \ac{DL} needs to be selected.
In order to determine in which mode each of the antennas should operate in, we define two binary
assignment vectors, $\vtX^u,\vtX^d\!\!\in\!\!\{0,1\}^{M\times 1}$, for \ac{UL} and \ac{DL},
respectively, such that
$x_{i}^{u} (x_{j}^{d})$ is equal to 1 if antenna $i$ ($j$) is used on \ac{UL} (\ac{DL}), or equal 
to 0 otherwise.
It is useful to transform the assignment vectors into diagonal assignment matrices, such
that $\mtX^u\!=\!\diag\left(\vtX^u\right)$ and $\mtX^d\!=\!\diag\left(\vtX^d\right) \in
\{0,1\}^{M\times M}$.
Consequently, we can apply $\mtX^u$ to the received \ac{UL} symbol $\vtY^u$, creating the
effective received symbol $\widetilde{\vtY}^u\!=\!\mtX^u\vtY^u\!\in\!\fdC^{M}$. Similarly, we can
apply $\mtX^d$ to the transmitted signal $\sum_{m=1}^{J} \vtW_m^ds_m^d \!+\! \vtC^d$, creating the
effective transmitted signal $\mtX^d\left(\sum_{m=1}^{J} \vtW_m^ds_m^d \!+\!
\vtC^d\right)\!\in\!\fdC^{M}$.

Using the antenna assignment, the signal models of~\eqref{eq:UL_signal}-\eqref{eq:DL_signal} can
be rewritten in a compact form as

\small\vspace*{-0.2cm}
\begin{align*}
\widetilde{\vtY}^u\! &=\! \sum\nolimits_{i=1}^{I} \widetilde{\vtH}^u_i(\sqrt{q_{i}^u}s_i^u+c_i^u) 
\!+\! \widetilde{\mtH}_{\text{SI}}\!\left(\sum\nolimits_{j=1}^J \vtW_j^d s_j^u \!+\! 
\vtC^d\!\right)\! \!+\! \widetilde{\vtEta}^u \!+\! \widetilde{\vtE}^u,\\ 
\widetilde{y}_j^d\! &=\! \widetilde{\vtH^u_i}^{\mathrm{H}}\!\left(\!\sum\nolimits_{m=1}^{J} \!
\vtW_m^ds_m^d \!+\! \vtC^d\!\right)\! \!+\! \sum\nolimits_{i=1}^I\! 
g_{ij}\!\left(\!\sqrt{q_{i}^u}s_i^u \!+\! c_i^u\right) \!+\! \eta^d_j \!+\! e^d_j, 
\end{align*}\normalsize

\noindent where $\widetilde{\vtH}^u_i\!\!=\!\!\mtX^u \vtH^u_i$, $\widetilde{\vtH}^d_j\!\!=\!\mtX^d 
\vtH^d_j$, $\widetilde{\mtH}_{\text{SI}}\!\!=\!\mtX^u \mtH_{\text{SI}} \mtX^d$ denote the
effective \ac{UL}, \ac{DL}, and \ac{SI} channels, respectively; $\widetilde{\vtEta}^u\!\!=\!\mtX^u 
\vtEta^u$ and $\widetilde{\vtE}^u\!\!=\! \mtX^u \vtE^u$ denote the effective noise and receiver 
distortion, with distributions $ \widetilde{\vtEta}^u 
\!\!\sim\!\stC\stN\left(\mathbf{0}_M,\sigma^2\mtX^u\right)$, and 
$\widetilde{\vtE}^u\!\!\sim\!\stC\stN\left(\mathbf{0}_M,\beta\mtX^u\diag(\mtPhi^u) \mtX^u\right)$, 
respectively.


We let the received signal, $\widetilde{\vtY}^u$, be linearly decoded at the \ac{BS} by a
filter $\vtR_i^u\in\fdC^{M\times 1}$.
Similarly, the received signal of \ac{DL} user $j$, $\widetilde{y}_j^d$, is linearly decoded by a
filter $r_j^d\in\fdC$. With this notation, we can write the \ac{MSE} of \ac{UL} user $i$ as 

\small\vspace*{-0.2cm}
\begin{align} 
\hspace*{-0.2cm}E_i^u = \bbE\Big\{\NormTwo{\Herm{\vtR_i^u}\widetilde{\vtY}^u-s_i^u}^2 \Big\}
 = \left\vert \sqrt{q_{i}^u}\Herm{\vtR_i^u} \widetilde{\vtH}_i^u -1 \right\vert^2
 \!+\!  \Herm{\vtR_i^u}\mtPsi_i^u\vtR_i^u.
\end{align}\normalsize
Similarly, let us define the \ac{MSE} of the received symbol of \ac{DL} user $j$ as 
\vspace*{-0.1cm}
\begin{align} 
\hspace*{-0.1cm}E_j^d \!=\! \bbE\Big\{\NormTwo{\Herm{r_j^d}\widetilde{y}^d_j-s_j^d}^2 \Big\} 
 \!=\! \left\vert \Herm{r_j^d} \widetilde{\vtH^d_j}^{\mathrm{H}}\vtW_j^d \!-\!1  \right\vert^2
 \!\!+\! \Abs{r_j^d}^2\Psi_j^d,
\end{align}
where $\mtPsi_i^u\!\in\!\fdC^{M\times M}$ and $\Psi_j^d\!\in\!\fdC$ are the covariance matrix and
variance of the total interference plus noise in the \ac{UL} and \ac{DL}, respectively, as
characterized by equations~\eqref{eq:Psi_UL}-\eqref{eq:Psi_DL}.
Notice that the expectations are taken with respect to the transmitted symbols and noise. In the 
above,

\small\vspace*{-0.4cm}
\begin{align}
\hspace*{-0.5cm} \mtPsi_i^u \!&\!=\! \sum\nolimits_{l\neq i}^{I}
q_l^u\widetilde{\vtH}_l^u\widetilde{\vtH^u_l}^{\mathrm{H}} \!+\!
\kappa \sum\nolimits_{l=1}^{I} q_l^u\widetilde{\vtH}_l^u\widetilde{\vtH^u_l}^{\mathrm{H}} \!+\!
\sum\nolimits_{j=1}^{J} \widetilde{\mtH}_{\text{SI}} \Bigg(\vtW_j^d\Herm{\vtW_j^d} 
\!+\!\nonumber\\
& \kappa\diag\Big(\vtW_j^d\Herm{\vtW_j^d}\Big) \Bigg)\Herm{\widetilde{\mtH}_{\text{SI}}} \!+\!
\beta\sum\nolimits_{l=1}^I q_l^u \diag\left( 
\widetilde{\vtH}_l^u\widetilde{\vtH^u_l}^{\mathrm{H}}\right) \!+\! \label{eq:Psi_UL}\nonumber\\
& \beta\sum\nolimits_{j=1}^{J}\diag\left(\widetilde{\mtH}_{\text{SI}} \vtW_j^d\Herm{\vtW_j^d}
\Herm{\widetilde{\mtH}_{\text{SI}}} \right) + \sigma^2\mtX^u,\\ 
\Psi_j^d \!&=\!\! \sum\nolimits_{m\neq j}^{J}\! \widetilde{\vtH^d_m}^{\mathrm{H}}
\vtW_m^d\Herm{\vtW_m^d} \widetilde{\vtH^d_m} \!+\! \kappa \sum\nolimits_{m= 1}^{J}\!
\widetilde{\vtH^d_m}^{\mathrm{H}}\! \diag\!\left(\! \vtW_m^d\Herm{\vtW_m^d}\!\right)
\widetilde{\vtH^d_m} \nonumber\\
&\hspace*{-0.2cm} \!+\! \sum\nolimits_{i=1}^{I} \Abs{g_{ij}}^2 q_i^u\left(\kappa \!+\! \beta 
\!+\! 1\right) \!+\! \beta \sum\nolimits_{m=1}^{J} \widetilde{\vtH^d_m}^{\mathrm{H}}
\vtW_m^d\Herm{\vtW_m^d}\widetilde{\vtH^d_m}  \!+\!\! \sigma^2 
\!\!.\hspace*{-0.5cm}\label{eq:Psi_DL}
\end{align}\normalsize

The optimal \ac{MSE} receiver can then be obtained by differentiating~\eqref{eq:obj_msemin} with
respect to either $\vtR_i^u$ or $r_j^d$, and setting it to zero. Notice that the derivatives are
taken with respect to complex numbers, and we therefore use the necessary definitions
from~\cite[Chapter 4]{Hjorungnes2011} to obtain the known \ac{MMSE} receivers for \ac{UL} and
\ac{DL} as

\small\vspace*{-0.3cm}
\begin{align}
\vtR_i^u &= \sqrt{q_{i}^u}\Inv{\left(q_i^u \widetilde{\vtH^u_i}\widetilde{\vtH^u_i}^{\mathrm{H}} +
\mtPsi_i^u\right)}\widetilde{\vtH^u_i},\label{eq:ul_opt_rec}\\
r_j^d &= \widetilde{\vtH^d_j}^{\mathrm{H}}\vtW_j^d
\Inv{\left(\widetilde{\vtH^d_j}^{\mathrm{H}}\vtW_j^d\Herm{\vtW_j^d}\widetilde{\vtH^d_j}
 + \Psi_j^d \right)},\label{eq:dl_opt_rec}
\end{align}\normalsize
where special care must be taken in the \ac{UL}, so that the inverse is taken disregarding
the zero columns/rows due to the usage of $\mtX^u$ and $\mtX^d$.
\vspace*{-0.1cm}
\subsection{Problem Formulation}\label{sub:prob_form}
Our goal is to minimize the sum \ac{MSE} of all users, with respect to \ac{UL}
and \ac{DL} antenna assignment.
We leverage the inverse relation to between user-rate and \ac{MSE}~\cite{Christensen2008}, to
formulate the antenna assignment problem as
\begin{subequations}\label{eq:sum_mse_prob}
\begin{align}
\underset{\substack{\vtX^u,\vtX^d,}}{\text{minimize}}\quad
& \sum\nolimits_{i=1}^I E_i^u + \sum\nolimits_{j=1}^J E_j^d \label{eq:obj_msemin}\\
\text{subject to}\quad & \vtX^u + \vtX^d = \vtOne,\label{eq:joint_ant}\\
\quad& \vtX^u,\vtX^d \in \{0,1\}^{M\times 1}, \label{eq:binary_x_uldl}
\end{align}
\end{subequations}
where constraint~\eqref{eq:joint_ant} ensures that antenna $k$ is used in either \ac{UL} or
\ac{DL} only.
To arrive at a solution of problem~\eqref{eq:sum_mse_prob}, we first rewrite it as a sum of two
quadratic and biquadratic terms.
We show that the problem is complex: provided that a solution for $\vtX^d$ (or
$\vtX^u$) is available, the problem is NP-hard in the remaining variable $\vtX^u$ ($\vtX^d$
respectively).
To circumvent the biquadratic term, we transform it into a quartic term and resort to a 
first-order approximation. 
To derive an approximate solution, we relax the binary
constraint~\eqref{eq:binary_x_uldl} to the unit hypercube, and use the framework of
\ac{PSCA}~\cite{Razaviyayn2014}.

\section{Solution Approach Based on \ac{PSCA}}\label{sec:scp_approx}

The sum \ac{MSE} problem~\eqref{eq:sum_mse_prob} can be expressed as the minimization of the sum 
of 
three functions: $f^u(\vtX^u)$, $f^d(\vtX^d)$, and $f^{u,d}(\vtX^u,\vtX^d)$, that
depend on $\vtX^u$, $\vtX^d$, and jointly on $\vtX^u$ and $\vtX^d$, respectively. 
Proposition~\ref{prop:quad_biquad_eq} poses this formulation, whose proof is in 
Appendix~\ref{app:proof_result}.
\vspace*{-0.2cm}
\begin{prop}\label{prop:quad_biquad_eq}
Consider optimization problem~\eqref{eq:sum_mse_prob}. Then, its objective function in 
~\eqref{eq:obj_msemin} can be written as the sum of two quadratic functions $f^u(\vtX^u)$, 
$f^d(\vtX^d)$ and one biquadratic function $f^{u,d}(\vtX^u,\vtX^d)$.
\end{prop}
\vspace*{-0.2cm}
Using Proposition~\ref{prop:quad_biquad_eq}, problem~\eqref{eq:sum_mse_prob} is equivalently 
stated as:
\begin{subequations}\label{eq:sum_f_ant}
\begin{align}
\underset{\vtX^u,\vtX^d}{\text{minimize}}\quad & f^u(\vtX^u) + f^{u,d}(\vtX^u,\vtX^d) + f^d(\vtX^d)
\label{eq:obj_f}\\
\text{subject to}\quad & \text{Constraints~\eqref{eq:joint_ant}-\eqref{eq:binary_x_uldl}}.
\end{align}
\end{subequations}
Notice that the optimal \ac{MMSE} filters in~\eqref{eq:ul_opt_rec}-\eqref{eq:dl_opt_rec} are used.
Moreover, for fixed $\vtX^d$(or $\vtX^u$), $f^{u,d}(\vtX^u,\vtX^d)$ is quadratic in $\vtX^u$ (or
$\vtX^d$).
It is well-known that the binary quadratic problem is NP-hard~\cite{Beck2000}, which implies that
the joint problem with variables $\{\vtX^u,\,\vtX^d\}$ and constraint~\eqref{eq:joint_ant} is
difficult to handle.

Initially, we can decouple $\vtX^u$ and $\vtX^d$ by considering that $\vtX^d = \vtOne - \vtX^u$.
Thus, we can write $f^u(\vtX^u)$ and $f^d(\vtX^d)$ as quadratic functions of $\vtX^u$. However,
$f^{u,d}(\vtX^u,\vtX^d)$ becomes a quartic function of $\vtX^u$. In view of making the problem
tractable, we consider a first-order Taylor approximation of $f^{u,d}(\vtX^u,\vtX^d)$ at a
neighbourhood of $\vtX^u$, denoted by $\widetilde{\vtX}^u$, as
\begin{equation}\label{eq:g_approx_func}
g^u(\vtX^u) = f^{u,d}(\widetilde{\mtX}^u) + \Trace{\Herm{\nabla f^{u,d}(\widetilde{\mtX}^u)
}(\mtX^u-\widetilde{\mtX}^u) },
\end{equation}
which is a linear function in $\vtX^u$, and recall that
$\widetilde{\mtX}^u\!=\!\diag\left(\widetilde{\vtX}^u\right)$. Using the derivative expressions of
$\Trace{\mtA\mtX\mtB}$~\cite[Chapter 4]{Hjorungnes2011} and applying the chain rule, the
gradient $\nabla f^{u,d} (\mtX^u)$ with respect to $\mtX^u$ is given by~\eqref{eq:grad_f_ud_xu}.
\begin{figure*}\footnotesize
\begin{align}\label{eq:grad_f_ud_xu}
\nabla\! f^{u,d}\! (\mtX^u)\! &\!=\!\!
\Bigg(\Transp{\mtR^u}\mtX^u\mtH_{\text{SI}}^*(\mtI_M-\mtX^u)
\Transp{\mtW^d}(\mtI_M-\mtX^u) \mtH_{\text{SI}}^{\text{T}} + \mtH_{\text{SI}}^*(\mtI_M-\mtX^u)
\Transp{\mtW^d}(\mtI_M-\mtX^u) \mtH_{\text{SI}}^{\text{T}}\mtX^u\Transp{\mtR^u} \nonumber\\
&\hspace*{0.4cm} + \diag\left(\beta\mtR^u\right)\mtX^u\mtH_{\text{SI}}^*(\mtI_M-\mtX^u)
\Transp{\mtSigma^d}(\mtI_M-\mtX^u)\mtH_{\text{SI}}^{\text{T}} + \mtH_{\text{SI}}^*(\mtI_M-\mtX^u)
\Transp{\mtSigma^d}(\mtI_M-\mtX^u)\mtH_{\text{SI}}^{\text{T}}\mtX^u\diag\left(\beta\mtR^u\right)
\Bigg)\nonumber\\
&\hspace*{0.4cm} -\Bigg( \mtH_{\text{SI}}^{\text{T}}\mtX^u\Transp{\mtR^u} \mtX^u
\mtH_{\text{SI}}^* (\mtI_M-\mtX^u) \Transp{\mtW^d} + \Transp{\mtW^d}(\mtI_M-\mtX^u)
\mtH_{\text{SI}}^{\text{T}}\mtX^u\Transp{\mtR^u} \mtX^u \mtH_{\text{SI}}^* +
\mtH_{\text{SI}}^{\text{T}}\mtX^u \diag\left(\beta\mtR^u\right) \mtX^u
\mtH_{\text{SI}}^*(\mtI_M-\mtX^u) \Transp{\mtSigma^d} \nonumber\\
&\hspace*{0.4cm}  + \Transp{\mtSigma^d}(\mtI_M-\mtX^u)
\mtH_{\text{SI}}^{\text{T}}\mtX^u \diag\left(\beta\mtR^u\right) \mtX^u \mtH_{\text{SI}}^*\Bigg).
\end{align}\normalsize
\hrulefill
\end{figure*}
Since we need not approximate $f^u(\vtX^u)$ and $f^d(\vtX^u)$, their respective approximations are
the functions themselves. Using Appendix~\ref{app:trace_diag_prop}, we can write the approximation
as a linear function of $\vtX^u$ as:
\begin{align}
g^u(\vtX^u)& = \Transp{\Diag{\nabla f^{u,d}(\widetilde{\mtX}^u)}}\vtX^u + c,
\end{align}
where $c$ is a constant that can be dropped from the optimization. Notice that the objective
function in problem~\eqref{eq:obj_f} becomes a quadratic function of $\vtX^u$.
Therefore, 
problem~\eqref{eq:obj_f} can be approximated as follows, where the objective
function approximates $g^u(\vtX^u)$ instead of $f^{u,d}(\vtX^u,\vtX^d)$:
\begin{subequations}\label{eq:sum_f_approx_ant}
\begin{align}
\underset{\vtX^u }{\text{minimize}}\quad & \Transp{\vtX^u}\mtLambda \vtX^u - 2 \Transp{\vtB} \vtX^u
\label{eq:non_hom_f_xu}\\
\text{subject to}\quad & \vtX^u \in \{0,1\}^{M\times 1},
\end{align}
\end{subequations}
where $\mtLambda\!\in\!\fdC^{M\times M}$ and $\vtB\!\in\!\fdR^M$ are denoted by
$\mtLambda \!=\!\mtLambda^u + \mtLambda^d $, and $\vtB \!=\! \vtA^u \!+\!
\Real{\!\mtLambda^d\vtOne\!} \!-\! \vtA^d \!-\! 0.5\Diag{\!\nabla
f^{u,d}(\widetilde{\mtX}^u) }\!$.

Due to the combinatorial nature of problem~\eqref{eq:sum_f_approx_ant}, we resort to a binary
relaxation, i.e., letting $\vtX^u\in [0,1]^M$. However, the approximation function
in~\eqref{eq:g_approx_func} holds for a neighbourhood of $\vtX^u$, and consequently requires an
iterative procedure to update the function approximation until convergence. With this, we use the
iterative convex approximation, \ac{PSCA} in~\cite{Razaviyayn2014}, where in addition to the
first-order approximation of the nonconvex function $f^{u,d}(\vtX^u,\vtX^d)$, the authors include
a proximal operator $0.5\alpha\NormTwo{\vtX^u - \vtX^{u^{(n)}}}^2$
to the objective function.
The reason for this choice is to find a compromise
between minimizing the function, and staying close to the previous iteration. Notice that
\ac{PSCA} is used for a single block of variables ($\vtX^u$), and does not apply the
parallelization allowed by the algorithm. With \ac{PSCA}, we update the point to the next
iteration using a constant step size rule $\rho$, which is proved to converge to a stationary
solution and has an iteration complexity of $\mathcal{O}(1/\epsilon)$~\cite[Theorem
3]{Razaviyayn2014}. With this, the optimization problem becomes
\begin{subequations}\label{eq:sum_f_approx_ant_rlx}
\begin{align}
\underset{\vtX^u}{\text{minimize}}\quad & \Transp{\vtX^u}\mtLambda \vtX^u - 2 \Transp{\vtB} \vtX^u
+ \frac{\alpha}{2}\NormTwo{\vtX^u - \vtX^{u^{(n)}} }^2
\label{eq:non_hom_f_xu_rlx}\\
\text{subject to} \quad & \vtX^u \in [0,1]^M.
\end{align}
\end{subequations}
Problem~\eqref{eq:sum_f_approx_ant_rlx} is convex and can be solved using well-known solvers, such
as CVX~\cite{CVX} or closed-form. Since the solution provided by \ac{PSCA} converges to a
stationary solution, we run it with $L$ differently chosen initial points, and choose the local
solution that provides the minimum \ac{MSE}.

Our proposed solution, termed RLX-PROX, is detailed in~\algref{alg:ant_assign_sol}.
It is centralized at the base station, and the inputs are the channel elements
$\vtH_{i}^u,\vtH_{j}^d,\mtH_{\text{SI}},g_{ij}$, the beamformers $\vtW_j^d$, \ac{UL} powers
$q_i^u$, and the optimization parameters $\epsilon,\alpha,\rho$. RLX-PROX randomly generates
the \ac{UL} antenna assignment $\vtX^u$, and starts the iterative \ac{PSCA}.
The iterations converge if the difference between the current and
the subsequent iterations are smaller than
$\epsilon$ (see line~\ref{alg_line:start_upd}). Subsequently, RLX-PROX evaluates the sum \ac{MSE}
and save the corresponding assignment (see
lines~\ref{alg_line:eval_sum_mse}-\ref{alg_line:save_x_iter}). 
After $L$ different
initializations, RLX-PROX selects the one that provides the minimum sum \ac{MSE}, and
performs the rounding to the binary set $\{0,1\}$ (see
lines~\ref{alg_line:sel_opt_x}-\ref{alg_line:output}).

\begin{algorithm}
 \footnotesize
  \caption{Approximated Solution RLX-PROX to
  Problem~\eqref{eq:sum_f_approx_ant}}\label{alg:ant_assign_sol}
 \begin{algorithmic}[1]
   \STATE Initialize
   $\vtH_{i}^u,\vtH_{j}^d,\mtH_{\text{SI}},g_{ij},\vtW_j^d,q_i^u,\epsilon,\alpha,\rho$
   \FOR{$l=1$ \TO $L$ }
     \STATE Initialize randomly $\vtX^u\in[0,1]^M$ such that $\vtX^u+\vtX^d=\vtOne$
     \STATE Set $n = 0$, $f_0(n)=0$, and $\widetilde{\vtX^u}^{(n)} = \vtX^u$
     \WHILE{ $ \Norm{\vtX^u - \vtX^{u^{(n)}} } > \epsilon$}\label{alg_line:start_upd}
         \STATE $n \gets n + 1$
         \STATE Update $\{\vtR_i^u(n), r_j^d(n)\}$ using
         equations~\eqref{eq:ul_opt_rec}-\eqref{eq:dl_opt_rec}\label{alg_line:eval_rx_filt}
         \STATE Solve problem~\eqref{eq:sum_f_approx_ant_rlx} to find
         $\widehat{\vtX^u}^{(n)}$\label{alg_line:eval_opt_prob}
         \STATE Update $\vtX^{u^{(n)}}$: $\vtX^{u^{(n)}} \gets
         \vtX^{u^{(n-1)}} +\rho \left(\widehat{\vtX^u}^{(n)} - \vtX^{u^{(n-1)}}\right)$
     \ENDWHILE
     \STATE Evaluate sum \ac{MSE}: $E^{u,d}_l \gets \sum\nolimits_{i=1}^I E_i^u +
     \sum\nolimits_{j=1}^J E_j^d$\label{alg_line:eval_sum_mse}
     \STATE Save assignment: $\vtX^u_l \gets \vtX^{u^{(n)}}$\label{alg_line:save_x_iter}
   \ENDFOR
   \STATE Select optimal across $L$ randomizations: $\vtX^{u^{\star}} \!\!=\!
   \argmin_{\vtX^u_l} E^{u,d}_l$ \label{alg_line:sel_opt_x} 
   \STATE Perform rounding to retrieve binary solution $\vtX^u$ from $\vtX^{u^{\star}}$
   \STATE \textbf{Output}: $\vtX^u$ as the approximate solution for
   problem~\eqref{eq:sum_mse_prob}\label{alg_line:output}
 \end{algorithmic}
\end{algorithm}

Therefore, algorithm~\ref{alg:ant_assign_sol} presents a centralized solution that can be employed 
on the time scale of large-scale fading.
The complexity of algorithm~\ref{alg:ant_assign_sol} is dominated by the matrix inversions (see
line~\ref{alg_line:eval_rx_filt}) and the solution of the optimization
problem~\eqref{eq:sum_f_approx_ant_rlx} (see line~\ref{alg_line:eval_opt_prob}), which has
$\mathcal{O}(M^3)$ worst-case complexity. Since we use $L$ randomizations, the overall worst-case
complexity is $\mathcal{O}(LM^3)$, which is much smaller than the complexity of the exhaustive
search solution $\mathcal{O}(2^M)$. 
\section{Numerical Results and Discussion}\label{sec:num_res_disc}
In this section we consider a single cell system operating in the pico cell
scenario~\cite{3gpp.36.828}.
The total number of antennas at the \ac{BS} varies from $M=8,\ldots,64$, and the total number of
served users is $I+J=8$, where we assume that $I=J$. The \ac{UL} transmission powers $q_i^u$ are
set to $P^u_{\text{max}}$, and the beamformers $\vtW_j^d$ are randomly generated with sum power
$P^d_{\text{max}}$ and fixed throughout the algorithm.
The parameters to obtain numerical results are listed in \tabref{tab:sim_param}.

\begin{table}
 \caption{Simulation parameters}\label{tab:sim_param}
 \scriptsize
 \begin{tabularx}{\columnwidth}{l|l}
    \hline 
    \textbf{Parameter}                         & \textbf{Value} \\
    \hline 
    Cell radius                                & \unit{40}{m} \\
    Number of \ac{UL}/\ac{DL} \acp{UE} $[I=J]$ & $4$  \\
    Monte Carlo iterations                     & $600$ \\ \hline
    Carrier frequency / System bandwidth       & \unit{2}{GHz}/\unit{10}{MHz} \\
    \ac{LOS}/\ac{NLOS} path-loss model         & Set according to~\cite[Table 6.2-1]{3gpp.36.828}\\
    Shadowing st. dev. \ac{LOS}/\ac{NLOS}      & \unit{3}{dB}/\unit{4}{dB}\\
    Thermal noise power $[\sigma^2]$           & \unit{-174.4}{dBm/Hz} \\
    Noise figure \ac{BS}/\ac{UL} user          & \unit{13}{dB}/\unit{9}{dB}\\
    Tx/Rx distortions $[\kappa\; \beta]$       & \unit{-120}{dB} (see~\cite{Day2012,Cirik2016})\\
    \ac{SI} cancelling level $\sigma_{\text{SI}}^2$ & $[-50\;\ldots\;-100]$~\unit{}{dB}\\ \hline
    \ac{BS}/\ac{UL} user maximum power $[P^d_{\text{max}}\; P^u_{\text{max}}]$ & $[30\;
    23]$~\unit{}{dBm}\\
    Optimization constants $[\epsilon\; \alpha \; \rho\; L]$ & $[10^{-3}\; 1\; 0.9\; 20]$\\
    \hline
\end{tabularx}
\end{table}

First, we benchmark our proposed algorithm, RLX-PROX, against and exhaustive search of
problem~\eqref{eq:sum_mse_prob}, termed EXH. 
We also compare these two solutions with a simple and common solution usually assumed in
full-duplex networks, based on equal splitting of the antennas between \ac{UL} and \ac{DL} with
$\vtX^u\!=\![\vtOne_{M/2}\; \vtZero_{M/2}]^{\mathrm{T}}$ and $\vtX^d\!=\![\vtZero_{M/2}\;
\vtOne_{M/2}]^{\mathrm{T}}$, referred to as SPLIT. In the following, we show the sum spectral
efficiency instead of the sum \ac{MSE} because of their inverse relation~\cite{Christensen2008}.

\begin{figure}
\centering
\includegraphics[width=0.8\linewidth,trim=0mm 0mm 0mm 0mm,clip]{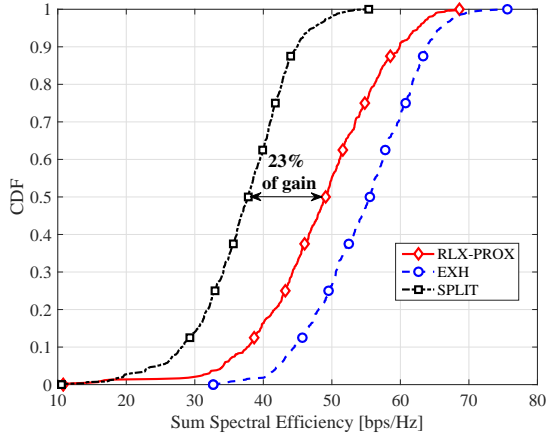}
\caption{\ac{CDF} of the sum spectral efficiency with \ac{SI} cancellation of \unit{-100}{dB}, and 
8 antennas at the \ac{BS}.
Notice that the optimality gap between our proposed RLX-PROX and EXH is small. Moreover, there is 
a substantial
gain of optimizing the antenna selection, since SPLIT is far from the other
solutions.}\label{fig:CDF_comp_Sum_SpEff}
\end{figure}

\figref{fig:CDF_comp_Sum_SpEff} shows the \ac{CDF} of the sum spectral efficiency of EXH, SPLIT
and our proposed RLX-PROX as a measure of the optimality gap, where a \ac{SI}
cancelling level of $\sigma_{\text{SI}}^2\!=$\unit{-100}{dB} is assumed.
Notice that the difference between EXH and RLX-PROX is small, where at the 50th percentile the
relative difference is approximately \unit{12}{\%}. Moreover, the gain provided by assigning the
antennas using the proposed RLX-PROX instead of the simple SPLIT is approximately \unit{23}{\%},
and of \unit{32}{\%} when compared to EXH.
\figref{fig:CDF_comp_Sum_SpEff} clearly shows that smart antenna assignment solutions for
\ac{UL} and \ac{DL} antennas provide substantial gains to full-duplex communications.

\begin{figure}
\centering
\includegraphics[width=0.8\linewidth,trim=0mm 0mm 0mm 0mm,clip]{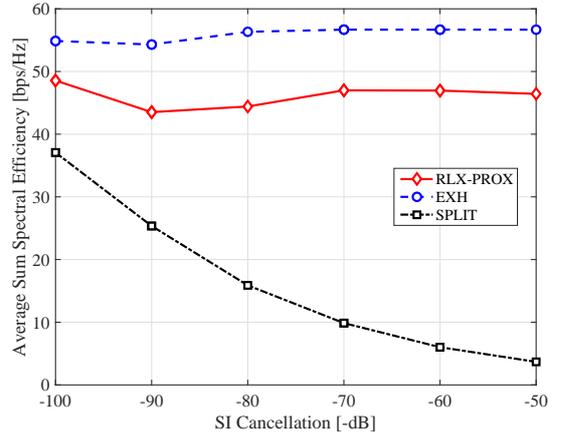}
\caption{Average sum spectral efficiency for different \ac{SI} cancellation capabilities, assuming 
8 antennas at the \ac{BS}. The optimality gap between EXH and our proposed RLX-PROX remains almost 
the same. On the contrary, the gap between the SPLIT and the other solutions only increases as the 
\ac{SI} cancellation capability decreases.}\label{fig:Avg_SI_Sum_SpEff}
\end{figure}

\figref{fig:Avg_SI_Sum_SpEff} shows the average sum spectral efficiency of EXH, SPLIT and
the proposed RLX-PROX, assuming different \ac{SI} cancelling levels, averaging over 600 Monte
Carlo iterations.
We observe that our proposed solution RLX-PROX maintains a high
average sum spectral efficiency and close to the optimal EXH for different \ac{SI} cancellation
capabilities. RLX-PROX and EXH maintain approximately the same average sum spectral efficiency for 
different \ac{SI} cancellation capabilities, with a slight increase with a reduced \ac{SI} 
capability. This behaviour is explained by the possibility of adapting the antenna assignment to 
the reduced \ac{SI} capability in RLX-PROX and EXH, which reduces the impact caused by the \ac{DL} 
users into \ac{UL} users.
Notice that the performance of SPLIT decreases as the \ac{SI} cancellation capability decreases, 
showing that antenna assignment between \ac{UL} and \ac{DL} is crucial for low \ac{SI} 
cancellation.

\begin{figure}
\centering
\includegraphics[width=0.8\linewidth,trim=0mm 0mm 0mm 0mm,clip]{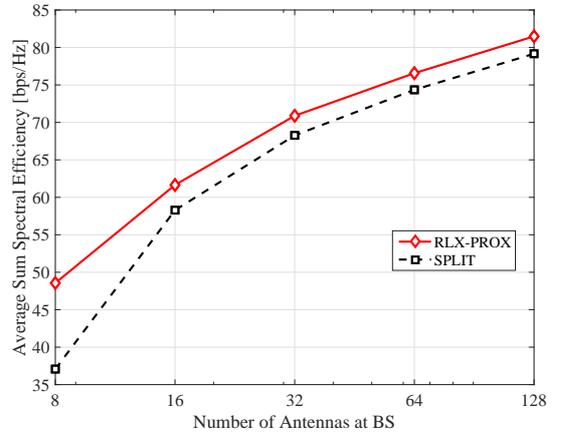}
\caption{Average sum spectral efficiency for different number of antennas and fixed number of
users. The gap between our proposed RLX-PROX and the SPLIT decreases with an increase in the
number of antennas.}\label{fig:Avg_Ant_Sum_SpEff}
\end{figure}

\figref{fig:Avg_Ant_Sum_SpEff} shows the average sum spectral efficiency of SPLIT and
the proposed RLX-PROX, assuming an increasing number of antennas at the \ac{BS}, fixed number of
users, and \ac{SI} cancelling level of $\sigma_{\text{SI}}^2=$\unit{-100}{dB}. The optimal 
solution EXH is not present because it has extremely high complexity with a large number of 
antennas.
Notice that the gap between RLX-PROX decreases when increasing the number of antennas.
For example, the gap decreases from approximately \unit{23}{\%} when using 8 antennas, to
approximately \unit{3}{\%} when using 128 antennas.
This behaviour implies that the role of antenna assignment is small when the number of antennas is
large.

Therefore, antenna assignment between \ac{UL} and \ac{DL} antennas (that is splitting UL/DL
antennas) provides substantial gains for
reduced number of antennas, for high and low \ac{SI} cancellation capabilities. When
the number of antennas is high, a simple antenna splitting is close to an optimized antenna
assignment.

\section{Conclusion}\label{sec:concl}
In this paper we considered the fundamental problem of splitting \ac{UL} and \ac{DL} antennas of a 
base station in full-duplex cellular communications. Specifically, our objective was to minimize 
the sum \ac{MSE} as a means of maximizing the sum spectral efficiency of \ac{UL} and \ac{DL} users.
This problem resulted in a binary nonlinear optimization problem, which is NP-hard. Due to its
combinatorial nature, we resorted to equivalent formulations,
\acl{PSCA}, 
and binary relaxation to obtain a stationary
solution. The proposed solution is guaranteed to converge to the relaxed problem with much smaller
complexity than exhaustive search.
The numerical results showed that the proposed solution improved the sum
spectral efficiency of a simple antenna splitting, while being close to the optimal exhaustive
search. We showed that antenna splitting is crucial when we have reduced number of
antennas, in which case our solution maintained a similar sum spectral efficiency across low and 
high \ac{SI} cancellation capabilities. For high number of antennas, a simple antenna splitting
achieved a performance close to our proposed solution, which showed that the role of antenna
splitting diminishes as the number of antennas increases.

For future works, we intend to study the impact of joint beamforming, power allocation along with
\ac{UL} and \ac{DL} antenna assignment using single- and multiple-antennas at the user side.
\appendices

\section{Proof of Proposition~\ref{prop:quad_biquad_eq}}\label{app:proof_result}
We define $f^u(\mtX^u)$ as
\small
\begin{align}
f^u(\mtX^u) \!&=\!\! \sum\nolimits_{i=1}^{I} \Big\{ \Herm{\vtR_i^u} \mtX^u \mtGamma_1^u
\mtX^u \vtR_i^u + \sigma^2 \Herm{\vtR_i^u} \mtX^u \vtR_i^u -\nonumber\\
&\hspace*{0.3cm}\!2\sqrt{q_i^u}\Real{\Herm{\vtR_i^u}\mtX^u\vtH^u_i} \Big\},
\end{align}\normalsize
where $\mtGamma_1^u\!\in\!\fdC^{M\times M}$ is defined as
\small
\begin{align}
\hspace*{-0.2cm}\mtGamma_1^u &\!=\! \sum\nolimits_{l=1}^{I} q_l^u \left\{ 
(\kappa+1)\vtH^u_l\Herm{\vtH^u_l} \!+\! \beta\diag\left(\vtH^u_l\Herm{\vtH^u_l}\right) \right\}.
\end{align}\normalsize
In addition, we let $f^d(\mtX^d)$ be

\small\vspace*{-0.3cm}
\begin{align}
\hspace*{-0.49cm}f^d(\mtX^d) &\!=\!\! \sum\nolimits_{j=1}^{J}\!\!\Bigg\{\! \Herm{\vtW_j^d} \mtX^d
\mtTheta^d_j \mtX^d
\vtW_j^d \!-\! 2\Real{\!\Herm{r_j^d}\Herm{\vtH_j^d}\mtX^d\vtW_j^d\!}
\!\Bigg\},
\end{align}\normalsize
where $\mtTheta_j^d\!\in\!\fdC^{M\times M}$ is defined as $\mtTheta_j^d \!\!=\! 
r^d(\beta+1)\mtH_j^d + r^d\kappa\diag\left(\mtH_j^d\right)$, with $r^d \!\!=\! \sum_{m=1}^{J}\! 
\Abs{r_m^d}^2$.
Finally, we denote $f^{u,d}(\mtX^u,\mtX^d)$ as

\small\vspace*{-0.4cm}
\begin{align}
\hspace*{-0.4cm}f^{u,d}(\mtX^u,\mtX^d) &\!=\!\! \sum\nolimits_{i=1}^{I}
\sum\nolimits_{j=1}^{J}\!\Herm{\vtR_i^u}
\Bigg\{\mtX^u\mtH_{\text{SI}}\mtX^d\Big(\vtW_j^d\Herm{\vtW_j^d} \!+\!\nonumber\\
& \!\kappa\diag\left(\vtW_j^d\Herm{\vtW_j^d} \right)\Big)\mtX^d\Herm{\mtH}_{\text{SI}}\mtX^u
+\nonumber\\
& \!\beta\diag
\Big(\mtX^u\mtH_{\text{SI}}\mtX^d\vtW_j^d\Herm{\vtW_j^d}\mtX^d\Herm{\mtH}_{\text{SI}}\mtX^u
 \Big) \!\Bigg\}\vtR_i^u.\!
\end{align}\normalsize

Using identities in Appendix~\ref{app:trace_diag_prop}, the above expressions can be rewritten in 
terms of $\vtX^u$ and $\vtX^d$ as
\begin{subequations}\label{eq:quad_f_ud}
\begin{align}
f^u(\vtX^u) &= \Transp{\vtX^u}\mtLambda^u\vtX^u -2\Transp{\vtA^u}\vtX^u,\\
f^d(\vtX^d) &= \Transp{\vtX^d}\mtLambda^d\vtX^d -2\Transp{\vtA^d}\vtX^d,
\end{align}
\end{subequations}
where the following matrices and vectors of dimensions $\fdC^{M\times M}$ and $\fdR^M$,
respectively, are defined as 
\small
\begin{align*}
\mtLambda^u &= \sum\nolimits_{i=1}^{I} \diag \left(\Herm{\vtR_i^u}\right) \mtGamma_1
\diag\left(\vtR_i^u\right),\\
\mtLambda^d &= \sum\nolimits_{j=1}^{J} \diag\left(\Herm{\vtW_j^d}\right) \mtTheta^d_j
\diag\left(\vtW_j^d\right),\\
\vtA^u \!&=\! \sum\nolimits_{i=1}^{I}\! \Bigg\{\! \sqrt{q_i^u} 
\Real{\!\Herm{\Diag{\vtR_i^u\Herm{\vtH_i^u} }} \!} \!-\! 
\frac{\sigma^2}{2}\Diag{\vtR_i^u\Herm{\vtR_i^u} }\! \Bigg\},\\
\vtA^d &= \sum\nolimits_{j=1}^{J} \Real{\Herm{r_j^d} \Diag{\vtH_j^d\Herm{\vtW_j^d} } }.
\end{align*}\normalsize
Similarly, we can write $f^{u,d}(\vtX^u,\vtX^d)$ as
\begin{align}\label{eq:biquad_f_ud}
\hspace*{-0.2cm}f^{u,d}(\vtX^u,\vtX^d) &=
\Trace{\mtX^u\mtH_{\text{SI}}\mtX^d\mtSigma^d\mtX^d\Herm{\mtH}_{\text{SI}}\mtX^u
\diag\left(\mtR^u\right)} + \nonumber\\
& \Trace{\mtX^u\mtH_{\text{SI}}\mtX^d\mtW^d\mtX^d\Herm{\mtH}_{\text{SI}}\mtX^u\mtR^u },
\end{align}
where we write $f^{u,d}(\vtX^u,\vtX^d)$ in terms of $\mtX^u,\mtX^d$ for simplicity, and define
matrices $\mtSigma^d,\mtR^u,\mtW^d\!\in\!\fdC^{M\times M}$ as 
$\mtSigma^d \!=\! \sum_{j=1}^{J}\! \vtW_j^d\Herm{\vtW_j^d}$, $\mtW^d \!=\! \mtSigma^d \!+\! 
\kappa\diag\left(\mtSigma^d\right)$, and $\mtR^u \!=\! \sum_{i=1}^{I} \vtR_i^u\Herm{\vtR_i^u}$, 
respectively. Notice that $f^{u,d}(\vtX^u,\vtX^d)$ is a biquadratic function of $\vtX^u$ and 
$\vtX^d$.
\vspace{-0.1cm}
\section{Useful Properties}\label{app:trace_diag_prop}
We enumerate some properties/identities used in the paper below, while omitting their
straightforward derivation:
\begin{enumerate}
\item \small $
\sum_{i}\Trace{\Herm{\vtX}\mtA_i\vtX}\!=\! \Trace{\left(\sum_{i}\mtA_i\right)\vtX\Herm{\vtX}};$
\item $\Trace{\diag(\vtX\Herm{\vtX})\mtA}\!=\!\Herm{\vtX}\diag(\mtA)\vtX;$
\item $ \Herm{\vtY}\diag\left(\vtX \right)\mtA\diag\left(\vtX \right)\vtZ =
\Herm{\vtX}\left(\Diag{\Herm{\vtY}}\mtA\;\Diag{\vtZ}\right)\vtX.$
\item $\Trace{\diag\left(\vtX^\star \right)\mtA\diag\left(\vtY \right)\Transp{\mtB} } =
\Herm{\vtX}\left(\mtA\odot\mtB \right)\vtY.$\normalsize
\end{enumerate}
\vspace{-0.2cm}

\bibliographystyle{IEEEtran}
\bibliography{FDref}

\end{document}